\documentclass[aps,prl,twocolumn]{revtex4}
\usepackage{amsmath}
\usepackage{amssymb}

\newcommand\ro{{\widehat\rho}}
\newcommand\dd{\mathrm{d}}
\newcommand\tr{\mathrm{tr}}
\newcommand\al{\alpha}
\newcommand\xo{\widehat x}

\newcommand\Oo{\widehat O}
\newcommand\zo{\widehat z}
\newcommand\s{\sigma}
\newcommand\T{\mathcal{T}}
\newcommand\M{\mathcal{M}}
\newcommand\N{\mathcal{N}}
\newcommand\e{\mathrm{e}}
\newcommand\half{{\scriptstyle\frac{1}{2}}}

\newcommand\Gt{\widetilde G}

\begin{document}
\title{Non-Markovian continuous quantum measurement of retarded observables}   
\author{Lajos Di\'osi}
\email{diosi@rmki.kfki.hu}
\homepage{www.rmki.kfki.hu/~diosi}
\affiliation{
Research Institute for Particle and Nuclear Physics\\
H-1525 Budapest 114, POB 49, Hungary}
\date{\today}

\begin{abstract}
We reconsider the non-Markovian time-continuous measurement of a Heisenberg
observable $\xo$ and show for the first time that it can be realized by an 
infinite set of entangled von Neumann detectors. The concept of continuous 
read-out is introduced and used to re-derive the non-Markovian stochastic 
Schr\"odinger equation. We can prove that, contrary to recent doubts, 
the resulting non-Markovian quantum trajectories are true single system trajectories
and correspond to the continuous measurement of a retarded functional of $\xo$.      
However, the generic non-Markovian trajectories are mixed state trajectories.

This version merges an Erratum [PRL, in print] with my Letter [PRL {\bf 100}, 080401 (2008)], 
some corrections follow directly from the criticism by Wiseman and Gambetta \cite{WisGam08}, 
further corrections restore the validity of my Letter. Contrary to my suggestion there,
the given continuous measurement schemes cannot yield \emph{pure state} trajectories but \emph{mixed-state}
ones \cite{WisGam08}. Yet, it is possible to retain my claim that the NMSE (5) describes true
time-continuous measurement - with \emph{delay} and \emph{retrodiction}.

\end{abstract}

\maketitle

Time-continuous measurement in quantum mechanics had long been an open theoretical issue 
because of the peculiarity of single quantum measurement itself. 
The Markovian theory emerged twenty years ago \cite{Gis84,Dio88,Bel88} from foundational
considerations. The requests in quantum optics (and elsewhere) triggered another, partly independent, line of
progress with expanding applications \cite{DalCasMol92plus}. So far the Markovian theory of continuous
measurement has become completely understood while the general non-Markovian one has remained 
an open issue even conceptionally.

Markovian time-continuous quantum measurement theory \cite{Dio88,Bel88} includes
the Markovian stochastic Schr\"odinger equation (MSSE) of the post-measurement 
state vector $\psi_t$, cf. \cite{Gis84}, as correlated with the read-out $x_t$ of 
the detector system that measures a certain Heisenberg observable $\xo_t$. 
A formal extension for the non-Markovian (even relativistic) 
case was published in ref.~\cite{Dio90}. This work calculated the asymptotic state $\psi_\infty[x]$ only, 
in function of the whole read-out $\{x_t;t\in(-\infty,\infty)\}$, and determined correctly 
its probability distribution functional $p_\infty[x]$. It could not interpret intermediate
conditional states because the concept of continuous read-out was missing.   
This incomplete non-Markovian continuous measurement theory remained largely ignored, 
it has not been improved or advanced. 
Meanwhile, Strunz found non-Markovian quantum trajectories \cite{Str96} and we invented their 
non-Markovian stochastic Schr\"odinger equation (NMSSE) \cite{DioStr97,DGS}. 
This NMSSE and its modifications have been studied in subsequent works 
\cite{SSEplus,GamWis0203,BasGhi02,AdlBas07}.  
Like in the Markovian case, one expected that
the solutions of the NMSSE turn out to be realizable on a single copy of our quantum system via infinite 
many von Neumann detectors coupled to it. Such realizability theorem holds for the solutions 
(quantum trajectories) of all diffusive MSSE \cite{WisDio01}. 
Yet, Gambetta and Wiseman conjectured that the solution of the NMSSE
can not be observed on a single system \cite{GamWis0203}; I wrote cautiously \cite{Dio06}: 
these non-Markovian trajectories can not be realized by any known 
way of monitoring \cite{Bre07}.

The present work reaches the positive conclusion: the non-Markovian trajectories are measurable single
system trajectories. A particular example can be the continuous measurement of a Heisenberg
coordinate $\xo_t$ with detectors of finite inertial time $1/\lambda$. 
Then the measured quantity becomes, e.g.:
\begin{equation}\label{zoretexp}
\zo_t=\lambda\int_0^t\e^{-\lambda(t-\s)}\xo_\s \dd \s~.
\end{equation}
Our work includes the more general case, see eq.~(\ref{zoret}) later. 
We describe the detector system and prove that the NMSSE is indeed the equation of the continuously measured state.    
The proofs are based on the approach of refs.~\cite{Dio90,Str96,DioStr97,DGS,SSEplus,GamWis0203}, 
an independent direct proof might be subject of future research. 
The knowledge of the superoperator formalism is a request; it can be understood 
from \cite{Dio90,Dio93} or learned from \cite{Choetal85}.

\emph{Stochastic unraveling.\/}
Assume that a Heisenberg variable $\xo_t$ of the
system couples for times $t\geq0$ to a harmonic reservoir variable whose 
equilibrium correlation function $\al(\tau-\s)$ will determine the reduced dynamics 
of the open system density operator $\ro_t$:
\begin{equation}\label{rot}
\ro_t=\M_t\ro_0~.
\end{equation}
For simplicity, let $\al(\tau-\s)$ be real.
Then the evolution superoperator $\M_t$ takes the following compact form \cite{Dio93}:
\begin{equation}\label{Mt}
\M_t=\T\exp\left(-\frac{1}{2}\int_0^t \dd\tau\int_0^t \dd\s \xo_{\tau,\Delta}\al(\tau-\s)\xo_{\s,\Delta}\right)~.
\end{equation}
Superoperator notation $\xo_{\tau,\Delta}\Oo$ means $[\xo_\tau,\Oo]$ for any
operator $\Oo$ standing to the right of $\xo_{\tau,\Delta}$ and
$\T$ prescribes time-ordering for all \emph{Heisenberg (super)operators} 
standing to the right of $\T$.

We could consider the reservoir as detector of $\xo_t$. Technically, it is more tractable if
we consider standard von Neumann detectors hence we replace the reservoir by them.
However, we require that their influence on the system be the same as the reservoir's. 
We assume, for simplicity, that the detectors are able to fully monitor the system's trajectory $\psi_t[x]$ for
all time $t\geq0$, in function of the detection read-out $\{x_\tau;\tau\in[0,t]\}$
whose probability distribution is denoted by $p_t[x]$. Then the stochastic mean of the trajectories will reproduce the
open system evolution:     
\begin{equation}\label{Mpsi_x}
\ro_t = \mathbf{M}\psi_t[x]\psi_t^\dagger[x]~,
\end{equation}
for all $t\geq0$, since the detector's influence is the same as the reservoir's. 
We say that the trajectories $\psi_t[x]$ \emph{unravel} the open system dynamics (\ref{rot}). 

In the Markovian special case $\al(\tau-\s)=g^2\delta(\tau-\s)$. 
Then the conditional state vector $\psi_t[x]$ satisfies the MSSE \cite{Gis84,Dio88,Bel88}.
The NMSSE \cite{DioStr97,DGS} became a candidate of being the equation of non-Markovian continuous
measurement of $\xo_t$. Here we use the simple real-noise version \cite{GamWis0203,BasGhi02,AdlBas07}.
For the unnormalized state vector $\Psi_t[z]$, the NMSSE reads:
\begin{equation}\label{SSE}
\frac{\dd\Psi_t[z]}{\dd t}=z_t\xo_t\Psi_t[z]-2\xo_t\int_0^t\!\!\!\!\al(t-\tau)\frac{\delta\Psi_t[z]}{\delta z_\tau}\dd\tau~,       
\end{equation}
where $z_\tau$ is a real random variable for $\tau\in[0,t]$.
The true post-measurement state is obtained via normalization 
$
\psi_t[z]=\Psi_t[z]/\Vert\Psi_t[z]\Vert~.       
$
The probability distribution of $z$ is the following: 
\begin{equation}\label{SSEptz}
p_t[z]=\Gt_{[0,t]}[z]~\Vert\Psi_t[z]\Vert^2~,
\end{equation}
where $\Gt_{[0,t]}[z]$ is defined by (\ref{Gt0tz}).
With this statistics, the solutions $\psi_t[z]$ unravel the non-Markovian open system dynamics (\ref{rot},\ref{Mt}):
\begin{equation}\label{Mpsi_z}
\ro_t = \mathbf{M}\psi_t[z]\psi_t^\dagger[z]~.
\end{equation}
Although to calculate the analytic form (\ref{SSEptz}) of $p_t[z]$ would be cumbersome, it follows from the method \cite{DGS} that
\begin{equation}\label{Mzt}
\mathbf{M}z_t=2\int_0^t\al(t-\s)\langle\xo_\s\rangle_t \dd \s~,
\end{equation}
where $\langle\xo_\s\rangle_t$ is $\xo_\s$'s quantum expectation value at time $t$ in the
conditional state $\psi_t[z]$.
This suggests that the NMSSE (\ref{SSE}) measures the retarded functional of $\xo_t$ rather
than $\xo_t$ itself. Compared to the Markovian case, 
there has been one serious issue left: Whether the trajectory $\psi_t[z]$ can, like the Markovian trajectories, be
realized on a single system by sequential von Neumann measurements of which $z_t$ is the read-out? 
We answer in the positive and construct the corresponding von Neumann detectors. 

\emph{Non-Markovian measurement device.\/}
The construction will be very similar to the Markovian one \cite{Dio88,Dio06Else} in that 
we replace the reservoir by a dense sequence of standard von Neumann detectors. 
To learn what happens, let us first consider a single von Neumann detector of initial density matrix $D_0(x;x')$ and
couple it to our system at time $\tau$ in order to measure the current Heisenberg operator $\xo_\tau$.
Following von Neumann (last three pages in \cite{Neu55}), 
we choose $\delta(t-\tau)\xo_\tau(-i\partial/\partial x)$ for the interaction Hamiltonian. 
We can write the initial composite state of the detector+system as $D_0(x;x')\ro_0$. 
Fortunately, we can and shall restrict all forthcoming calculations on the elements $x=x'$ 
since we shall eventually collapse on (or trace over) the pointer coordinates. 
After the interaction, the total state  becomes entangled at $\tau$ and the pointer $x$ 
gets shifted by $\xo_\tau$:
\begin{equation}\label{vNshiftx}
D_0(x;x)\ro_0\longrightarrow D_0(x-\xo_{\tau,L}; x-\xo_{\tau,R})\ro_0~.
\end{equation}
In superoperator notations $\xo_{\tau,L}\Oo=\xo_\tau\Oo$ and $\xo_{\tau,R}\Oo=\Oo\xo_\tau$.
It is the \emph{read-out} of the pointer $x$ that turns the total state into the following conditional
post-measurement state, depending on the read-out, of the system alone:
\begin{equation}\label{vNrox}
\ro(x)=\frac{1}{p(x)}D_0(x-\xo_{\tau,L};~x-\xo_{\tau,R})\ro_0~.
\end{equation}
The read-out $x$ has the probability distribution $p(x)$ whose expression follows from the
normalization of the above conditional state:
\begin{equation}\label{vNpx}
p(x)=\tr D_0(x-\xo_{\tau,L};~x-\xo_{\tau,R})\ro_0~.
\end{equation}
  
Now, let us choose a fine discretization $\tau=n\epsilon$ of the time, 
$n=0,\pm1,\pm2,\dots$.
We install an infinite sequence of von Neumann detectors, they could be numbered by the integers $n$
but we label them by the discretized times $\tau=n\epsilon$.   
The pointer coordinates of the detectors will be respectively denoted by $x_\tau$.
The detector of \emph{label} $\tau=n\epsilon$ measures the Heisenberg operator $\xo_\tau$ of the system
via the mechanism (\ref{vNshiftx}-\ref{vNpx}) provided we switch the von Neumann interactions on. 
We do so for the non-negative \emph{labels}, i.e., we choose the interaction Hamiltonian
$\sum_{\tau\geq0}\delta(t-\tau)\xo_\tau(-i\partial/\partial x_\tau)$.

We depart from the Markovian construction and assume \emph{initially correlated detectors}.
Let their initial wave function be:
\begin{equation}
\phi_0[x]=\sqrt{\N}\exp\left(-\epsilon^2\sum_{\tau,\s}x_\tau\al(\tau-\s)x_\s\right)~,
\end{equation}
where the summation extends for all discretized values of both $\tau$ and $\s$. 
The notation $[x]$ anticipates the continuous (or weak measurement) limit \cite{Dio88,Dio06Else} $\epsilon\rightarrow0$ 
where the above wave function becomes the square root of the Gaussian functional (\ref{Gx}), i.e.: $\phi_0[x]=\sqrt{G[x]}$.
We carry out the explanation in the continuous limit. The total initial density matrix reads:
\begin{equation}
\ro_0[x;x']=\sqrt{G[x]}\ro_0\sqrt{G[x']}~.
\end{equation}
As we switched on the detectors of labels $\tau\geq0$ only, 
at time $t>0$ each pointer coordinate $x_\tau$ with $\tau\in[0,t]$ will have been shifted     
by $\xo_\tau$ and the following composite state emerges [cf.~(\ref{vNshiftx})]:
\begin{equation}\label{}
\ro_t[x;x]=
\T \sqrt{G[x-\theta_{[0,t]}\xo_L]}\sqrt{G[x-\theta_{[0,t]}\xo_R]}\ro_0~,
\end{equation}
where $\theta_{[0,t]}$ denotes the characteristic function $\theta_{[0,t]}(\tau)$ of the period $[0,t]$.
This can be written into the following compact form:
\begin{equation}\label{rotxx}
\ro_t[x;x]=\T G[x-\theta_{[0,t]}\xo_c]\M_t\ro_0~,
\end{equation}
using the eqs.~(\ref{Mt},\ref{Gx}) and the superoperator notation $\xo_c\Oo=\half\{\xo,\Oo\}$.
This remarkable novel form guarantees explicitly that the
reduced density matrix $\ro_t$ of the system satisfies the open system evolution (\ref{rot},\ref{Mt}) as
it should. Indeed, the tracing over the detectors' Hilbert space is equivalent with the 
functional integration of the diagonal elements (\ref{rotxx}) over all $x_\tau$, which cancels the factor 
$G$ and leaves us with  (\ref{rot}). 

\emph{Continuous read-out.\/}
It is crucial to realize that the true time-evolution of the system's conditional state 
depends on our chosen schedule of reading out the pointers $x_\tau$. 
\emph{We can read out any $x_\tau$ at any time} 
since all detectors are always available. 
Of course, we better read out the value $x_\tau$ at a \emph{time} which is later than the \emph{label} $\tau$
of the detector because the detector will only have coupled to the system at time $\tau$. Hence,
a natural schedule is that we read out $x_\tau$ immediately at time $\tau$. Hence,
until any given time $t>0$ we would read out all pointers $x_\tau$ for the period $[0,t]$ and no
others. To calculate the conditional post-measurement state $\ro_t[x]$ of the system at time $t$, we trace (integrate) 
the total density matrix (\ref{rotxx}) over all $x_\tau$ with $\tau\notin[0,t]$: 
\begin{equation}\label{rotxrotxx}
\ro_t[x]=\frac{1}{p_t[x]}\int \ro_t[x;x] \prod_{\tau\notin[0,t]}\dd x_\tau~.
\end{equation}
This post-measurement density matrix $\ro_t[x]$ of the system depends on the  
read-outs $x_\tau$ of $\tau$ from $[0,t]$ only.
By substituting (\ref{rotxx}),
we obtain:
\begin{equation}\label{rotx}
\ro_t[x]=\frac{1}{p_t[x]}\T G_{[0,t]}[x-\xo_c]\M_t\ro_0~,
\end{equation}
where $G_{[0,t]}[x]$ is the marginal distribution of $G[x]$, similarly to (\ref{Gt0tz}).
This is our ultimate equation for the non-Markovian continuous measurement of the Heisenberg
observable $\xo_t$, completing the theory \cite{Dio90} (which only gave $\ro_\infty[x]$). 
Recall that, as always, the denominator $p_t[x]$ assures $\tr\ro_t[x]=1$ 
as well as it yields the probability distribution of the read-outs. Contrary to our
assumption, the state (17) is not pure even if it started from a pure $\ro_0$; 
the continuous readout of $x_t$ cannot provide sufficient information for a
pure state $\psi_t[x]$, as shown by Wiseman and Gambetta \cite{WisGam08}.

In order to find the measurement process that corresponds to the NMSSE (\ref{SSE}), 
we alter our read-out schedule. Instead of the Heisenberg variables $\{x_\tau;\tau\in[0,t]\}$ 
we read out the following linear functional of them: 
\begin{equation}
z_\tau=2\int_{-\infty}^\infty \al(\tau-\s)x_\s \dd \s~,
\end{equation}
which we also write as $z=2\al x$. We re-express the total density matrix (\ref{rotxx})
in the new pointer variables:
\begin{equation}\label{rotzz}
\ro_t[z;z]=\T\Gt[z-2\al\theta_{[0,t]}\xo_c]\M_t\ro_0~,
\end{equation}
where we used the identity $G[x]=\mathrm{Jacobian}\times\Gt[z]$. 
Again, we suppose that we read out each pointer of \emph{label} $\tau$
(i.e.: $z_\tau$) at \emph{time} $\tau$. 
Until time $t>0$, this schedule implies that
all pointers $z_\tau$ for the period $[0,t]$ are read out and the rest of them are not.
The conditional state of the system is defined by: 
\begin{equation}\label{rotzrotzz}
\ro_t[z]=\frac{1}{p_t[z]}\int \ro_t[z;z] \prod_{\tau\notin[0,t]}\dd z_\tau~,
\end{equation}
which transforms (\ref{rotzz}) into:
\begin{equation}\label{rotz}
\ro_t[z]=\frac{1}{p_t[z]}\T\Gt_{[0,t]}[z-2\al\theta_{[0,t]}\xo_c]\M_t\ro_0~,
\end{equation}
where $\Gt_{[0,t]}[z]$ is the marginal distribution (\ref{Gt0tz}) of $\Gt[z]$.
This is our ultimate equation for the non-Markovian continuous measurement of the
observable 
\begin{equation}\label{zoret}
\zo_t=2\int_0^t \al(t-\s)\xo_\s \dd \s
\end{equation}
which is a \emph{retarded} functional of the Heisenberg variable $\xo_\tau$.
This interpretation of $\ro_t[z]$ can shortly be inspected. 
Recall that \emph{at time $t$} we read out the pointer \emph{of label $t$}, i.e.: $z_t$. 
The factor $\Gt_{[0,t]}[z-2\al\theta_{[0,t]}\xo_c]$
in the expression (\ref{rotz}) of the measured state shows that at time $t$ the pointer 
$z_t$ localizes around (i.e.: measures) the observable (\ref{zoret}). The eq.~(\ref{Mzt}) 
holds between the read-out $z_t$ in (\ref{rotz}) and the retarded variable $\zo_t$ (\ref{zoret});
instead of the direct proof we are going to prove the complete equivalence of the NMSSE (\ref{SSE})
with our construction summarized by eq.~(\ref{rotz}).

\emph{Stochastic Schr\"odinger Equation.\/}
We are going to prove that the NMSSE (\ref{SSE}) governs the evolution (\ref{rotz}). 
Let us find $\ro_t[z]$ in the form:
\begin{equation}\label{rotzPsi}
\ro_t[z]=\frac{1}{p_t[z]}\Gt_{[0,t]}[z]\Psi_t[z]\Psi_t^\dagger[z]~,
\end{equation}
where $\Psi_t[z]$ is the unnormalized conditional state vector of the system.
Taking the trace of both sides, the norm condition yields exactly the $p_t[z]$ (\ref{SSEptz}) that belongs
to the NMSSE (\ref{SSE}).
Inserting (\ref{rotzPsi}) as well as $\ro_0=\psi_0\psi_0^\dagger$ into (\ref{rotz}), 
it reduces to:
\begin{equation}
\Psi_t[z]\Psi_t^\dagger[z]=
\frac{1}{\Gt_{[0,t]}[z]}\T\Gt_{[0,t]}[z-2\al\theta_{[0,t]}\xo_c]\M_t\psi_0\psi_0^\dagger~.
\end{equation}
Substituting eqs.~(\ref{Mt}) and (\ref{Gt0tz}), the r.h.s. factorizes and we can write equivalently:
\begin{equation}
\Psi_t[z]=\!
\T\!\!\exp\!\left(\int_0^t\!\!\!\! z_\tau \xo_\tau \dd \tau
     -\!\!\int_0^t\!\!\!\!\dd\tau\!\!\int_0^t\!\!\!\!\dd\s \xo_\tau \al(\tau-\s)\xo_\s\!\!\right)\!\psi_0~.
\end{equation}
This $\Psi_t[z]$ is the solution of the NMSSE (\ref{SSE}), as can be seen by substitution. 
That completes our proof.

\emph{Delayed continuous readout.\/}
Unfortunately, the chosen readout schedule alters the reduced dynamics (2) because the detector modes (18)
are not retarded, hence the coupling between the system and the detector mode $z_\tau$ 
continues after time $\tau$, cf. Ref.~\cite{WisGam08}. 
It ceases, nonetheless, at $\tau+T$ provided $T>0$ is much larger than the reservoir correlation time so that 
$\al(T)=0$ be already a good approximation. We can thus keep the reduced dynamics (2) invariant if
we apply continuous readout with a finite \emph{delay} $T$. We read out each pointer of \emph{label} 
$\tau$ (i.e.: $z_\tau$) at \emph{time} $\tau+T$. The conditional state (20) of the system at time $t>T$ 
must be replaced by:
\begin{equation}\label{rotz_del}
\ro_t[z;\mathrm{delay}\!\!=\!\!T]=\frac{1}{p_t[z;\mathrm{delay}\!\!=\!\!T]}
\int\ro_t[z;z]\!\!\!\!\!\!\prod_{\tau\notin[0,t-T]}\!\!\!\!\!\!\dd z_\tau~.
\end{equation}
It turns out that $p_t[z;\mathrm{delay}\!\!=\!\!T]=p_{t-T}[z]$, i.e., 
the statistics of delayed continuous readouts obtained until time $t$ is identical
to the statistics of zero-delay (and all-in-one \cite{WisGam08}) readouts until time $t-T$. 
The delayed-readout state obviously coincides with the following average of the zero-delay-readout states (20): 
\begin{eqnarray}\label{rotz_del_rotz}
\ro_t[z;\mathrm{delay}\!\!=\!\!T]
&=&\frac{1}{p_{t-T}[z]}\int\ro_t[z]~p_t[z]\!\!\!\!\!\!\prod_{\tau\in[t-T,t]}\!\!\!\!\!\!\dd z_\tau\\
&=&\frac{1}{p_{t-T}[z]}\int\psi_t[z]\psi_t^\dagger[z]~p_t[z]\!\!\!\!\!\!\prod_{\tau\in[t-T,t]}\!\!\!\!\!\!\dd z_\tau~.
\nonumber
\end{eqnarray}
The second equality follows from the insertion of $\ro_t[z]=\psi_t[z]\psi_t^\dagger[z]$ where, according to 
Eq.~(23), the pure state $\psi_t[z]$ must be the normalized solution $\psi_t[z]=\Psi_t[z]/\Vert\Psi_t[z]\Vert$ of the NMSE (5). 
As we see, $\psi_t[z]$ does not directly describe a continuously measured quantum trajectory because the values 
$z_\tau$ for $\tau\in[t-T,t]$ would belong to the all-in-one measurement at time $t$. Still, the above partial average of 
the pure states $\psi_t[z]$ over those fictitious $z_\tau$ does fully describe our (delayed) non-Markovian 
continuous measurement. The normalized solutions $\psi_t[z]$ of the NMSE (5) do correspond to \emph{retrodicted} pure states of
the system, the proof and physical interpretation will be given elsewhere.

\emph{Summary.\/}
We proved for the first time that both the formalism \cite{Dio90} of non-Markovian measurement theory  and 
the NMSSE \cite{DioStr97} are equivalent with using correlated von Neumann detectors in 
the weak-measurement continuous limit, i.e., with the continuous read-out of the values of a given retarded 
functional of a Heisenberg variable on a single quantum system. Our merit is the constructive
proof of existence of the underlying standard quantum mechanical measurement process.
The results should be generalized in various directions. 
We can interpret complex reservoir correlation functions, too, if we include the mechanism of feed-back 
\cite{Dio90}. We might retain the original reservoir as   
detector \cite{GamWis0203}, to extract information by measuring the reservoir but without altering the 
non-Markovian reduced dynamics of the monitored system. Then the measured retarded observable might be 
identified by a reservoir field. (Theories advocating non-Markovian 
stochastic \emph{modification} of quantum theory \cite{BasGhi02,AdlBas07,Pea99plus} refuse the 
measurement interpretation of the stochastic field.) 
The concept of relativistically invariant continuous measurement \cite{Dio90} can be reconsidered
for the intermediate states $\psi_t[x]$ as well.  
Our work might lead to efficient numeric simulation algorithms or,
conversely, might make us understand why they don't exist.

\emph{Appendix.--}
Let $x_\tau$ be a random time-dependent real variable and consider the normalized 
Gaussian distribution functional of $\{x_\tau;\tau\in(-\infty,\infty)\}$:
\begin{equation}\label{Gx}
G[x]=\N\exp\left(-2\int_{-\infty}^\infty\!\!\!\!\dd\tau\!\!\int_{-\infty}^\infty\!\!\!\!\dd\s x_\tau\al(\tau-\s)x_\s\right)~,
\end{equation}
$\al(\tau-\s)$ is a real positive definite kernel, We define its inverse by
$\int_{-\infty}^\infty \al^{-1}(\tau-s)\al(s-\s) \dd s = \delta(\tau-\s)$.
Introduce the normalized functional Fourier transform of $G[x]$, too:
\begin{equation}\label{Gtz}
\Gt[z]=\widetilde\N\exp
\left(-\frac{1}{2}\int_{-\infty}^\infty\!\!\!\!\dd\tau\!\!\int_{-\infty}^\infty\!\!\!\!\dd\s z_\tau\al^{-1}(\tau-\s)z_\s\right)~.
\end{equation}
We need certain marginal distributions as well, e.g.: 
\begin{equation}\label{Gt0tz}
\Gt_{[0,t]}[z]=\int \Gt[z] \prod_{\tau\notin[0,t]} \dd z_\tau~, 
\end{equation}
and similarly for $G_{[0,t]}[x]$.
These marginal distributions are also Gaussian, e.g.: 
\begin{equation}\label{Gt0tzal}
\Gt_{[0,t]}[z]=\widetilde\N_{[0,t]}\exp\!
\left(\!-\frac{1}{2}\!\int_0^t\!\!\!\!\dd\tau \!\int_0^t\!\!\!\!\dd\s z_\tau\al^{-1}_{[0,t]}(\tau,\s)z_\s\!\right)~,
\end{equation}
where the restricted new kernel $\al^{-1}_{[0,t]}(\tau,\s)$ is defined by
$\int_0^t \al^{-1}_{[0,t]}(\tau,s)\al(s-\s) \dd s = \delta(\tau-\s)$ for all $\tau,\s\in[0,t]$.
 
This work was supported by the
Hungarian Scientific Research Fund under Grant No 49384.

\end{document}